\IfFileExists{aastex702.cls}{%
  \documentclass[twocolumn,times,tighten,twocolappendix]{aastex702}%
}{%
  \documentclass[twocolumn,times,tighten,twocolappendix]{aastex631}%
}

\usepackage[T1]{fontenc}
\usepackage{graphicx}
\usepackage{amsmath}
\usepackage{booktabs}
\usepackage{array}
\usepackage{xcolor}
\usepackage{url}
\usepackage{hyperref}

\graphicspath{{./}}
\providecommand{\uat}[2]{#1 (#2)}

\shorttitle{Aperture Photometric Accuracy of PSF-deconvolved Images}
\shortauthors{Zhang et al.}

\received{2026 April 28}
\revised{2026 May 27}
\accepted{2026 June 1}
\published{2026 July 1}
\journalinfo{The Astronomical Journal, 172:67 (18pp), 2026 July\hfill https://doi.org/10.3847/1538-3881/ae7718}
\makeatletter
\renewcommand{\frontmatter@title@above}{%
\vspace*{-0.55in}%
\footnotesize\noindent\textsc{\@journalinfo}\par
\vspace{0.15in}%
}
\makeatother

\begin{document}

\title{Aperture Photometric Accuracy of Point-spread-function-deconvolved Astronomical Images}

\author{Shen Zhang}
\affiliation{National Astronomical Observatories, Chinese Academy of Sciences, Beijing 100101, People's Republic of China}
\affiliation{School of Astronomy and Space Science, University of Chinese Academy of Sciences, Beijing 100049, People's Republic of China}
\author{Lei Wang}
\affiliation{Purple Mountain Observatory, Chinese Academy of Sciences, Nanjing 210023, People's Republic of China; leiwang@pmo.ac.cn}
\email[show]{leiwang@pmo.ac.cn}
\author{Yachong Diao}
\affiliation{Beijing City University, No. 26 Muyan Road (Yangzhen Section), Yangzhen, Shunyi District, Beijing 101309, People's Republic of China}
\author{Zhaojun Yan}
\affiliation{Shanghai Astronomical Observatory, Chinese Academy of Sciences, Shanghai 200030, People's Republic of China}
\author{Xiyan Peng}
\affiliation{Shanghai Astronomical Observatory, Chinese Academy of Sciences, Shanghai 200030, People's Republic of China}
\author{Yao Liu}
\affiliation{School of Physical Science and Technology, Southwest Jiaotong University, Chengdu 610031, People's Republic of China}
\author{Huanyuan Shan}
\affiliation{Shanghai Astronomical Observatory, Chinese Academy of Sciences, Shanghai 200030, People's Republic of China}
\affiliation{School of Astronomy and Space Science, University of Chinese Academy of Sciences, Beijing 100049, People's Republic of China}
\author{Guanghuan Wang}
\affiliation{Purple Mountain Observatory, Chinese Academy of Sciences, Nanjing 210023, People's Republic of China}
\author{Fengshan Liu}
\affiliation{National Astronomical Observatories, Chinese Academy of Sciences, Beijing 100101, People's Republic of China}
\affiliation{School of Astronomy and Space Science, University of Chinese Academy of Sciences, Beijing 100049, People's Republic of China}
\author{Chengliang Wei}
\affiliation{Purple Mountain Observatory, Chinese Academy of Sciences, Nanjing 210023, People's Republic of China}
\author{Lin Nie}
\affiliation{Shanghai Astronomical Observatory, Chinese Academy of Sciences, Shanghai 200030, People's Republic of China}

\author{Xuechun Chen}
\affiliation{Beijing City University, No. 26 Muyan Road (Yangzhen Section), Yangzhen, Shunyi District, Beijing 101309, People's Republic of China}
\author{Hengkai Ding}
\affiliation{Purple Mountain Observatory, Chinese Academy of Sciences, Nanjing 210023, People's Republic of China}
\author{Zhenya Zheng}
\affiliation{Shanghai Astronomical Observatory, Chinese Academy of Sciences, Shanghai 200030, People's Republic of China}
\affiliation{School of Astronomy and Space Science, University of Chinese Academy of Sciences, Beijing 100049, People's Republic of China}

\makeatletter
\@namedef{email@i}{}      
\@namedef{email@iii}{}    
\@namedef{email@iv}{}     
\@namedef{email@v}{}      
\@namedef{email@vi}{}     
\@namedef{email@vii}{}    
\@namedef{email@viii}{}   
\@namedef{email@ix}{}     
\@namedef{email@x}{}      
\@namedef{email@xi}{}     
\@namedef{email@xii}{}    
\@namedef{email@xiii}{}   
\@namedef{email@xiv}{}    
\makeatother

\begin{abstract}
This paper introduces an improved Richardson--Lucy (RL) framework utilizing negative backfilling (NB) and the novel filtered image deconvolution and noise backfilling (FiDNoBa) algorithm. By integrating a low-pass prefiltering stage---specifically employing a Kalman filter---with the reintegration of filtered-out noise components after RL iterations, we effectively suppress noise artifacts while maintaining rigorous flux integrity. We extensively evaluate our Kalman + RL + NB approach using both numerical simulations and real observational data. Simulations across three regimes demonstrate superior photometric accuracy, reducing the relative mean deviation (RMD) to below 1\% for isolated sources, maintaining high fidelity (RMD $\sim$0.37\%) for blended S\'{e}rsic galaxies, and achieving $\sim$5\% RMD in crowded stellar fields while significantly increasing source recovery rates compared to traditional point-spread function photometry. Crucially, we validate the practical efficacy of FiDNoBa on real space-based (JWST/NIRCam) and ground-based (Very Large Telescope Survey Telescope OmegaCAM) images. The results confirm that our method reliably handles complex instrumental signatures, such as the extended wings of the STPSF, mitigating the severe ringing artifacts typical of traditional RL deconvolution while ensuring highly stable aperture photometry in practical survey processing.
\end{abstract}

\keywords{\uat{Astronomy data analysis}{1858} --- \uat{Astronomy image processing}{2306} --- \uat{Deconvolution}{1910}}

\section{Introduction}
\label{sec1}
The point-spread function (PSF) describes the response of an optical system to a point source. Consequently, any image captured by an optical system is equivalent to the convolution of the true image (at infinite resolution) with the PSF. Convolution with the PSF smooths the image, erases high-frequency details, and limits the spatial resolution to the diffraction limit of the system (e.g., the Rayleigh criterion). Based on the mathematical principles governing the impact of the PSF on image resolution, numerous image deconvolution algorithms have been proposed (for a comprehensive review, see \citealp{Starck2002}).

Subsequently, image deconvolution algorithms have been widely applied in medical imaging (e.g., \citealp{Shajkofci2020}) and astronomy (e.g., \citealp{Farrens2017,Wang2022}). In regimes with high signal-to-noise ratios (SNRs), these algorithms can significantly improve the achieved spatial resolution, enabling the reliable recovery and accurate estimation of the intrinsic properties of celestial sources (\citealp{McNeil2005,Leist2024}). In astronomy, several key challenges remain unresolved, including the detection of faint sources, photometry of bright extended sources, measurement of active galactic nucleus (AGN) light curves, light-curve analysis of gravitationally lensed quasars, and photometry in crowded stellar fields. Researchers have actively explored these problems using PSF deconvolution algorithms (e.g., \citealp{Millon2024,Wang2025}).

From a mathematical perspective, the imaging process of an astronomical instrument can be described by the convolution equation at coordinates $(x, y)$ (\citealt{Bracewell1954}):
\begin{equation}
I(x,y) = (O \ast P)(x,y) + N(x,y)
\end{equation}
where $I(x,y)$ denotes the calibrated observed image, $O(x,y)$ is the ``true'' image before convolution, $P(x,y)$ represents the instrumental PSF, $N(x,y)$ is the noise introduced during the observation (e.g., detector noise, optical system noise, and sky background noise), and $*$ denotes the convolution operation. Deconvolution is the inverse process of convolution, aiming to recover the true image $O(x,y)$ from the observed image $I(x,y)$ and the PSF $P(x,y)$ \citep{Starck2002}. In an ideal scenario, deconvolution would perfectly restore the true image $O(x,y)$. In practice, however, the presence of high-spatial-frequency noise -- such as noise with a spatial profile sharper than the convolution kernel (PSF), including Poisson, Gaussian, and salt-and-pepper noise -- renders the deconvolution problem ill posed, meaning a stable solution cannot be obtained. Researchers therefore rely on regularization methods to find approximate solutions.

Through simulations of astronomical images, \cite{McNeil2005} concluded that the Richardson--Lucy (RL) algorithm (\citealp{Richardson1972,Lucy1974}), which is based on maximum likelihood analysis, yields the best deconvolution results for point sources (e.g., stars), particularly in terms of flux conservation. Using MIRISim simulations of JWST/MIRI multiband observations of simplified AGN models, \cite{Leist2024} found that the Kraken algorithm (\citealp{Hope2022}) minimizes the FWHM of the nuclear source while maintaining good photometric accuracy across all simulated bands. With their chosen iteration stopping criterion, Kraken exhibited a clear advantage in visual quality; however, the authors acknowledged that Kraken is inferior to the myopic iterative step-preserving algorithm AIDA (\citealp{Mugnier2004}) and the RL algorithm in terms of flux conservation, and does not significantly outperform other algorithms in recovering morphological parameters (position angle and eccentricity) of extended sources. By decomposing the deconvolved image into point-source and extended-source channels, \cite{Millon2024} proposed the STARRED method, which can measure AGN light curves and gravitationally lensed quasar light curves. This algorithm requires stacking multiepoch exposure images and combining them with PSF photometry, achieving promising results in simulations. STARRED relies on the degree of decoupling between the point-source and extended-source channels; when the extended and point sources are strongly coupled across all epochs, it cannot accurately measure point-source variability. Nevertheless, its deconvolution principle shares similarities with the RL algorithm, as both maximize a likelihood function but differ in their regularization strategies.

The RL algorithm iteratively seeks an approximate solution using the following formula:
\begin{equation}
O^{(k+1)} = O^{(k)} \cdot \left( P^{T} * \frac{I}{P * O^{(k)}} \right)
\end{equation}
where $P^{T}$ is the transpose (or adjoint) of the PSF. The RL algorithm is a robust linear regularized iterative method widely used for image deconvolution in astronomy and other disciplines. Its key advantages are (1) each iteration yields a nonnegative result; (2) when the background noise is greater than zero, the flux of the image array is conserved both globally and locally; and (3) as it is derived from a Poisson noise model, it is naturally suited for astronomical images. However, it also has well-documented drawbacks: (1) noise amplification; (2) the emergence of ringing artifacts at high spatial frequencies, both of which worsen with increasing iteration number (\citealt{Magain1998}); and (3) the inability to directly deconvolve images containing negative values (e.g., background-subtracted images). Given the widespread applicability and robust reliability of the RL algorithm, this study adopts it for all PSF deconvolution calculations, focusing on mitigating its noise amplification and enabling the handling of negative-valued images, followed by an evaluation of the resulting aperture photometry accuracy.

In practical astronomical image processing, the use of the RL algorithm often leads to dilemmas: (1) Performing RL deconvolution on images with a positive mean background causes flux nonconservation -- the algorithm erroneously assigns background flux to the source and produces ringing (Gibbs pattern) around bright sources; (2) Subtracting the background (setting the mean background to zero) introduces negative values, which violate the nonnegativity condition required for RL deconvolution. To resolve these dilemmas, \cite{Shepp1982} proposed an improved RL algorithm that includes a background term, adding an expected background term in the denominator of the original formula:
\begin{equation}
O^{(k+1)} = O^{(k)} \cdot \left( P^{T} * \frac{I}{P * O^{(k)} + b} \right)
\end{equation}
where $b$ is the expected background \citep{Shepp1982}.

Another approach is the nonnegative constraint, i.e., setting all negative values (mostly noise) to zero, as negative photon counts are physically impossible. This constraint is automatically enforced by the multiplicative update mechanism of the algorithm, ensuring the physical plausibility of the image \citep{Prato2024}. However, in practice, we often need to process background-subtracted images; this approach retains only positive noise, effectively introducing a positive bias, and flux conservation remains problematic.

In this paper, we first improve the nonnegative constraint method by introducing negative noise backfilling, enabling the RL-deconvolved image to better preserve flux conservation. Second, by supplementing with local low-pass filtering and filtered-out noise backfilling, we further improve the flux conservation of the RL algorithm and suppress its noise amplification to a certain extent. Finally, we apply these new techniques to the PSF deconvolution of simulated point sources, extended sources (galaxies), and crowded stellar fields, achieving more accurate aperture photometry than traditional methods, and in some cases attaining precision comparable to PSF-fitting photometry. The paper is organized as follows. Section~\ref{sec2} describes the principle of negative backfilling and the initial toy mock test. Section~\ref{sec3} presents the filtered image deconvolution and noise backfilling (FiDNoBa) algorithm and its constituent low-pass filtering components. Section~\ref{sec4} details the numerical simulation results across three distinct regimes: isolated point sources (Mock-I), close S\'{e}rsic galaxies (Mock-II), and crowded stellar fields (Mock-III). Section~\ref{sec5} demonstrates the practical application of our framework to real observational data from both space-based (JWST/NIRCam) and ground-based (VST-OmegaCAM) telescopes. Finally, Section~\ref{sec6} provides our conclusions.

\section{negative backfilling and Toy Mock Test}
\label{sec2}
Through a series of simulations, we found that a nonzero background is the source of flux nonconservation in RL-deconvolved images, and it also causes ringing around bright sources, which undermines flux conservation in aperture photometry. Therefore, ensuring a zero-mean background is a prerequisite for flux conservation in RL deconvolution. We add one step to the traditional nonnegative constraint method: negative noise backfilling.

\begin{enumerate}
    \item Separate the zero-mean image into a negative part (mostly noise) and a nonnegative part.
    \item Apply RL deconvolution to the nonnegative part.
    \item After each iteration, add back the previously separated negative noise image.
\end{enumerate}

Negative noise backfilling ensures the flux conservation of the RL-deconvolved image for aperture photometry. In order to facilitate the comparison of various algorithms, we choose the number of RL deconvolution iterations such that the curve of growth (COG, or encircled energy as a function of radius) of the flux on the deconvolved image becomes horizontal or reaches the actual input flux, which occurs within 1 $\sigma$ of the deconvolution Gaussian kernel.

We designed a toy mock test (Mock-0) using an isolated point source. The point source has an input flux of 6000 counts, which was convolved with a PSF (FWHM = 8 pixels) to obtain a blurred image. A constant background of 10 counts was then added, the entire image was Poissonized, and finally, the constant background of 10 counts was subtracted to obtain the mock image ($63\times63$ pixels). By varying the Poisson random seed, we generated 200 realizations (i.e., 200 mock images). We applied three methods to these mock images: the {\bf original RL} (adding back the constant background before deconvolution and subtracting it after deconvolution), the {\bf improved RL}\citep{Shepp1982}, and the NB RL, each with 200 RL iterations.

\begin{figure}
\centering
\includegraphics[width=\columnwidth]{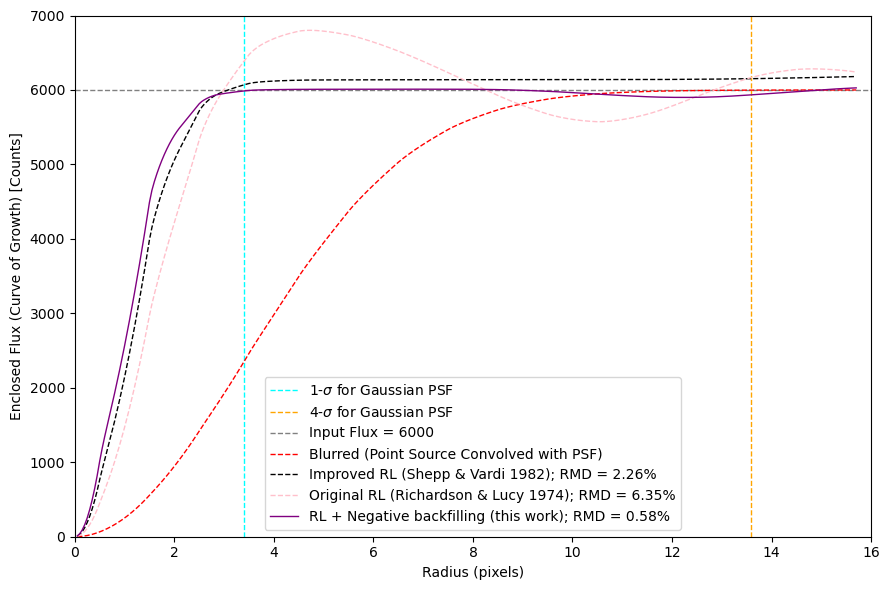}
\caption{COG centered on the point source for the toy mock. Gray dashed line: input flux; red dashed line: cumulative flux of the blurred image; pink dashed line: original RL; black dashed line: improved RL; and purple solid line: NB. The A14 region where the relative mean deviation (RMD) statistic is evaluated lies between the cyan and orange dashed lines.}
\label{fig1}
\end{figure}

Figure \ref{fig1} shows the COG for the toy mock, averaged over 100 mock images and 200 RL iterations. The curve for the blurred image (red dashed line) is clearly extended. After PSF deconvolution, the growth curves become very steep in the inner region (e.g., radius $<2$ pixels in this example) and generally flat in the outer region. Notably, the improved RL (black dashed line) and NB RL (purple solid line) exhibit a long plateau near 6000 counts, indicating that the total flux within that region is highly insensitive to the choice of aperture radius. The original RL (pink dashed line) struggles to conserve flux, and its oscillatory shape suggests strong ringing artifacts when deconvolving an image with a background.

We define the annular region from $1\sigma$ to $4\sigma$ as the A14 region, where $\sigma$ is the standard deviation of the input Gaussian PSF. Furthermore, we define the relative mean deviation (RMD) of the flux growth curve within the A14 region as a metric of flux conservation. The RMD values are reported in the caption of Figure \ref{fig1}. The RMD metric shows that the improved RL algorithm already outperforms the original RL significantly, but still has a systematic bias of $\sim2\%$, consistently overestimating the true flux within the A14 region. Negative backfilling achieves a bias of $\sim0.6\%$, an improvement of a factor of $\sim3$ over the improved RL, and it exhibits a long, almost horizontal high plateau within the A14 region, which is beneficial for the robustness of aperture photometry.

\section{Filtered-out Noise Backfilling Algorithm}
\label{sec3}

Through the minor improvement of the nonnegative constraint method (negative backfilling (NB) and the toy mock test, we uncovered a rule for ensuring flux conservation. On the other hand, it is well known that the RL algorithm amplifies noise. For example, before negative backfilling, the RL deconvolution of the nonnegative part inevitably amplifies positive noise \citep{Starck2002,Wang2022}. This is undesirable; instead, we aim for the RL algorithm to deconvolve only the signal component. Combining these two insights, we further propose a ``surgical'' solution: the FiDNoBa algorithm.

\begin{enumerate}
    \item Apply a small-kernel low-pass filter (e.g., $3\times3$ pixels) to the background-subtracted image $A$ (which contains negative values and noise) to obtain a filtered image $B$ and the filtered-out noise $A-B$.
    \item Separate the zero-mean filtered image $B$ into a negative part $B_1$ and a nonnegative part $B_0$.
    \item Apply RL deconvolution to the nonnegative part $B_0$ to obtain $C_0$, then add back the negative part to obtain $C = C_0 + B_1$.
    \item Finally, add the previously separated filtered-out noise to the synthesized image $C$ to obtain the final image $D = C + (A-B)$.
\end{enumerate}

Backfilling the filtered-out noise largely recovers the flux lost during filtering, while the small-kernel low-pass filter avoids excessive smoothing of spatial details and targets only high-frequency noise. To test the effects of different types of low-pass filters, we examined five commonly used filters: Kalman, mean, median, BM3D, and NLM (Non-Local Means).

{\bf Kalman Filter} is a foundational recursive algorithm for optimal state estimation in dynamic linear systems subject to Gaussian noise. Its primary strength lies in its ability to efficiently fuse prior system models with noisy real-time measurements to produce a minimum mean-square error estimate with low computational overhead. However, the filter's efficacy is strictly contingent upon the linearity of the system dynamics and the Gaussian nature of the noise distribution, often requiring complex extensions (e.g., Extended or Unscented Kalman Filters) for non-linear applications. This seminal approach was originally introduced by \cite{Kalman1960}.

{\bf Mean Filter} (box filter) is the most fundamental linear spatial smoothing technique, operating by replacing each pixel value with the arithmetic average of its neighbors. While its principal advantage is exceptional computational simplicity and speed -- making it suitable for real-time preprocessing -- this comes at the cost of significant image degradation. Specifically, the uniform weighting of all pixels within the kernel inevitably leads to the blurring of critical edge features and fails to distinguish between noise and signal, rendering it ineffective against impulse noise.

{\bf Median Filter} is a non-linear operator designed specifically to suppress impulse noise (e.g., salt-and-pepper artifacts) while preserving edge integrity. By sorting pixel values within a local window and selecting the median, the filter effectively eliminates outlier values without the smearing effect characteristic of averaging methods. Nevertheless, this robustness against extreme values can result in the erosion of fine details, thin lines, and sharp corners when the filtering window is large or the noise density is high, and it offers suboptimal performance for Gaussian noise reduction.

{\bf BM3D} (Block-Matching and 3D Filtering) is a state-of-the-art traditional method for image denoising, leveraging the non-local self-similarity of natural images to achieve superior restoration quality. Its core innovation involves grouping similar image patches into 3D arrays and applying collaborative filtering in a transform domain, which allows for exceptional noise suppression while meticulously preserving textures and structural details. The primary drawback of BM3D, however, is its substantial computational complexity, arising from the exhaustive block-matching search and multi-stage 3D transformations, which often precludes its use in real-time systems without hardware acceleration. This method was formally detailed by \cite{Dabov2007}.

{\bf NLM} (Non-Local Means) revolutionized image denoising by extending the concept of averaging from local neighborhoods to the entire image based on patch similarity. By assigning weights to pixels according to the geometric configuration of their surrounding patches rather than just spatial proximity, NLM excels at retaining complex textures and repetitive patterns that local filters typically smooth out. Despite its theoretical elegance and visual superiority in texture preservation, the algorithm suffers from prohibitive computational costs due to its quadratic complexity $\mathcal{O}(N^2)$, which requires comparing every pixel patch against a vast search window. This approach was first proposed by \cite{Buades2005}.

\begin{figure}
\centering
\includegraphics[width=\columnwidth]{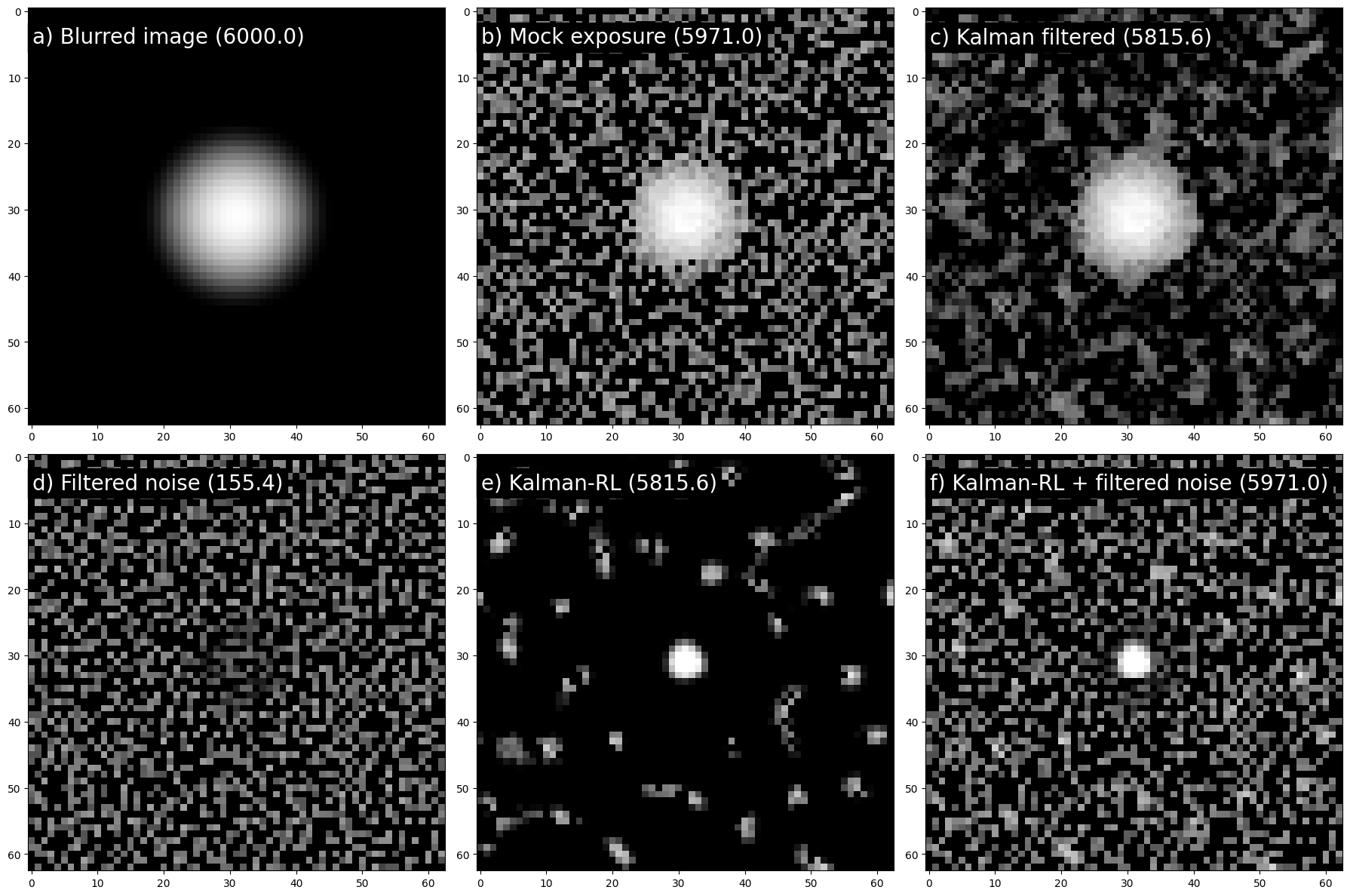}
\caption{Demonstration of the FiDNoBa algorithm using the Kalman filter. Numbers in parentheses are the total flux (counts). Panel (c): image after Kalman filtering; (d): filtered-out noise = panel (b) $-$ panel (c); (e): Kalman-filtered RL-deconvolved image (nonnegative part of panel (c) deconvolved, then negative part added back); and (f): final result after noise backfilling (panel (e) + (d)).}
\label{fig2}
\end{figure}

\section{Mock Tests}
\label{sec4}

\subsection{Mock-I: Isolated Point Sources}

We performed a simulation similar to the previous toy mock (Mock-I), but with the addition of the five filters (Kalman, mean, median, BM3D, NLM) to implement the FiDNoBa algorithm. Figure \ref{fig2} shows a demonstration of the FiDNoBa algorithm using the Kalman filter as an example; the numbers in parentheses represent the total flux (counts). After convolving the point source with a Gaussian PSF\footnote{ For ground-based telescopes, a Gaussian PSF is more representative compared to a PSF with pronounced azimuthal structure.}, we obtained Panel $({\bold a})$. Adding a constant background of 10 counts and Poissonizing, followed by subtracting the constant background of 10 counts, yielded Panel $({\bold b})$. Panel $({\bold c})$ shows the image after Kalman filtering; Panel $({\bold d})$ represents the filtered-out noise, i.e., (b$-$c). By taking the nonnegative part of the Kalman-filtered image $({\bold c})$, applying RL deconvolution, and then adding back the negative part of $({\bold c})$, we obtained the Kalman-filtered RL-deconvolved image $({\bold e})$. Panel $({\bold f})$ shows the result after noise backfilling, i.e., adding the noise from Panel $({\bold d})$ to the result $({\bold e})$.

\begin{figure*}
\centering
\includegraphics[width=0.9\textwidth]{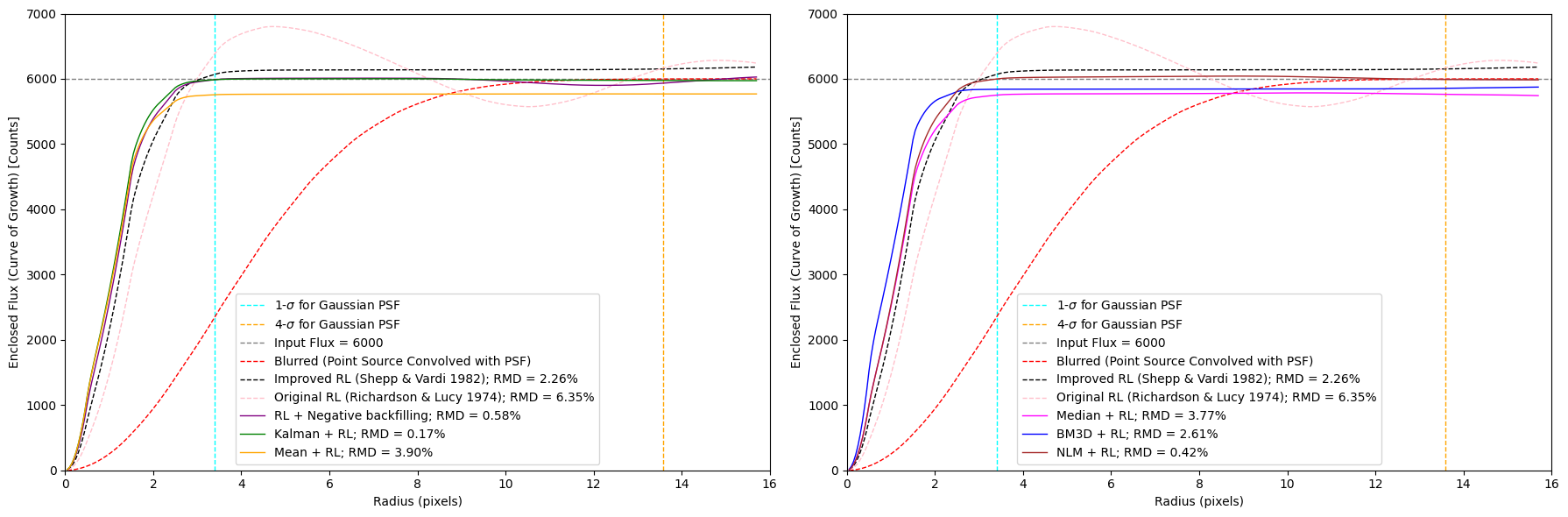}
\includegraphics[width=0.9\textwidth]{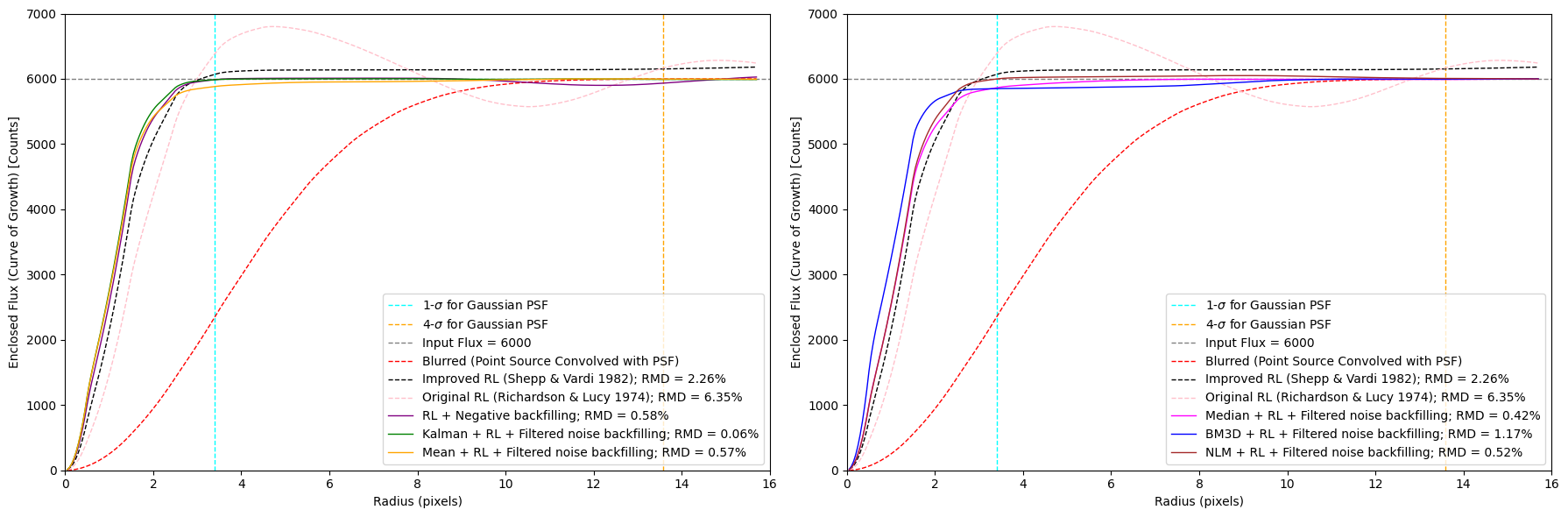}
\caption{Growth curves after different filters and RL deconvolution, without (upper row) and with (lower row) noise backfilling. Input flux = 6000 counts, background = 10 counts, and FWHM = 8 pixels. All RL runs performed 200 iterations, and the curves are averages over 200 realizations.}
\label{fig3}
\end{figure*}

To thoroughly compare FiDNoBa with traditional algorithms, we varied the input flux over three levels (2000, 6000, 18000 counts), the background noise over two levels (10 and 50 counts), and the Gaussian PSF FWHM over two values (4 and 8 pixels). For each filter, we considered cases with and without noise backfilling. All simulation images and PSFs were $63\times63$ pixels, the number of RL iterations was fixed at 200, and for each configuration, we generated 200 independent realizations by varying the Poisson random seed. Figure \ref{fig3} shows the growth curves for the case of input flux 6000, background 10, and FWHM = 8 pixels, comparing the cases without noise backfilling (upper row) and with noise backfilling (lower row). Dashed lines represent traditional algorithms, while solid lines represent algorithms with filtering or noise backfilling. Comparing the upper and lower rows, it is clear that the FiDNoBa algorithms (filtered RL + noise backfilling) generally have a significant advantage in flux conservation (except for the NLM algorithm, which is also the sole exception in Table \ref{tab:rmd_comparison}).

A broader comparison is presented in Table \ref{tab:rmd_comparison}, where, for example, ``F2kW4B50'' denotes input flux 2000, FWHM 4 pixels, and background 50. From Table \ref{tab:rmd_comparison}, we observe that most algorithms with noise backfilling are significantly better at flux conservation than those without noise backfilling. The Kalman-filtered RL images have better flux conservation than the unfiltered RL+NB, and with the addition of noise backfilling, Kalman+RL+NB achieves the best performance among all algorithms (values in bold).

\begin{table*}
  \centering
  \caption{Flux Conservation Relative Mean Deviation Metric for Various Algorithms in Mock-I}
  \label{tab:rmd_comparison}
  \resizebox{\textwidth}{!}{
  \begin{tabular}{lcccccc}
    \toprule
    RMD in percent & F2kW4B50 & F2kW8B10 & F6kW4B50 & F6kW8B10 & F18kW4B50 & F18kW8B10 \\
    \midrule
    improved RL     & 4.73 & 3.85 & 2.47 & 2.26 & 1.15 & 1.16 \\
    original RL     & 9.63 & 9.44 & 6.62 & 6.35 & 4.10 & 3.91 \\
    RL+NB           & 0.98 & 1.72 & 0.26 & 0.58 & 0.09 & 0.20 \\
    Kalman+RL       & 0.87 & 1.49 & 0.30 & 0.17 & 0.08 & 0.07 \\
    {\bf Kalman+RL+NB}    & {\bf 0.61} & {\bf 0.85} & {\bf 0.23} & {\bf 0.06} & {\bf 0.06} & {\bf 0.05} \\
    Mean+RL         & 4.50 & 9.45 & 2.17 & 3.90 & 1.39 & 1.61 \\
    Mean+RL+NB      & 1.46 & 2.39 & 2.42 & 0.57 & 2.83 & 0.27 \\
    Median+RL       & 12.10& 5.22 & 12.61& 3.77 & 12.70& 3.31 \\
    Median+RL+NB    & 1.79 & 1.31 & 1.38 & 0.42 & 1.11 & 0.32 \\
    BM3D+RL         & 13.70& 8.26 & 4.86 & 2.61 & 2.07 & 1.01 \\
    BM3D+RL+NB      & 4.62 & 3.14 & 2.14 & 1.17 & 1.09 & 0.55 \\
    NLM+RL          & 2.19 & 13.47& 0.33 & 0.42 & 0.08 & 0.22 \\
    NLM+RL+NB       & 0.89 & 3.34 & 0.33 & 0.52 & 0.08 & 0.22 \\
    \bottomrule
  \end{tabular}
  }
\end{table*}

\subsection{Mock-II: Close S\'ersic Galaxies}

\begin{figure}
\centering
\includegraphics[width=\columnwidth]{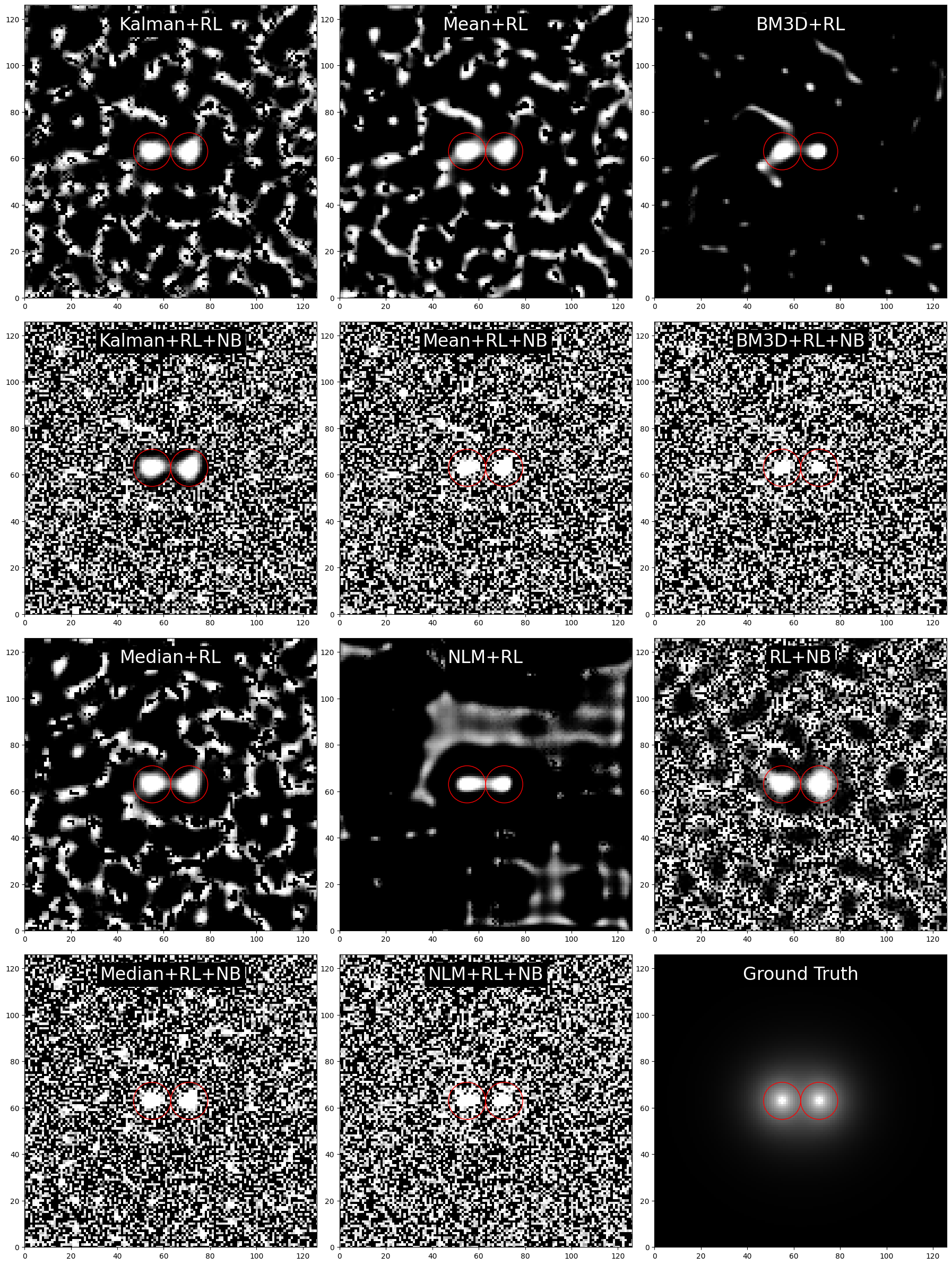}
\caption{Simulated images of two closely separated, equally bright S\'{e}rsic-profile galaxies (extended sources) with a center-to-center separation of 16 pixels, a total flux of 6000 counts per source, and a S\'{e}rsic index $n=4.0$ (ground truth, lower right panel). The ground truth is convolved with a Gaussian PSF (FWHM = 8 pixels), a background of 50 counts is added, and the image is Poissonized to generate the simulated exposure image (Mock-II). Red circular apertures with a radius of 8 pixels mark the photometric regions after RL deconvolution (100 iterations) using six methods: Kalman, mean, median, BM3D, NLM, and RL+NB.}
\label{fig4}
\end{figure}

We simulated two very close, equally bright extended sources (galaxies) using S\'ersic profiles, with a center separation of 16 pixels. Each source has a total flux of 6000 counts and a S\'ersic index $n=4.0$, as shown in the lower right panel of Figure \ref{fig4} (ground truth). The ground truth was convolved with a Gaussian PSF (FWHM = 8 pixels), a background of 50 counts was added, and the image was Poissonized to produce a simulated exposure image; this constitutes Mock-II. We processed this mock with six methods (Kalman, mean, median, BM3D, NLM, and RL+NB) followed by RL deconvolution (100 iterations), and then performed aperture photometry using a circular aperture of radius 8 pixels (red circles in Figure \ref{fig4}).

\begin{figure*}
\centering
\includegraphics[width=0.9\textwidth]{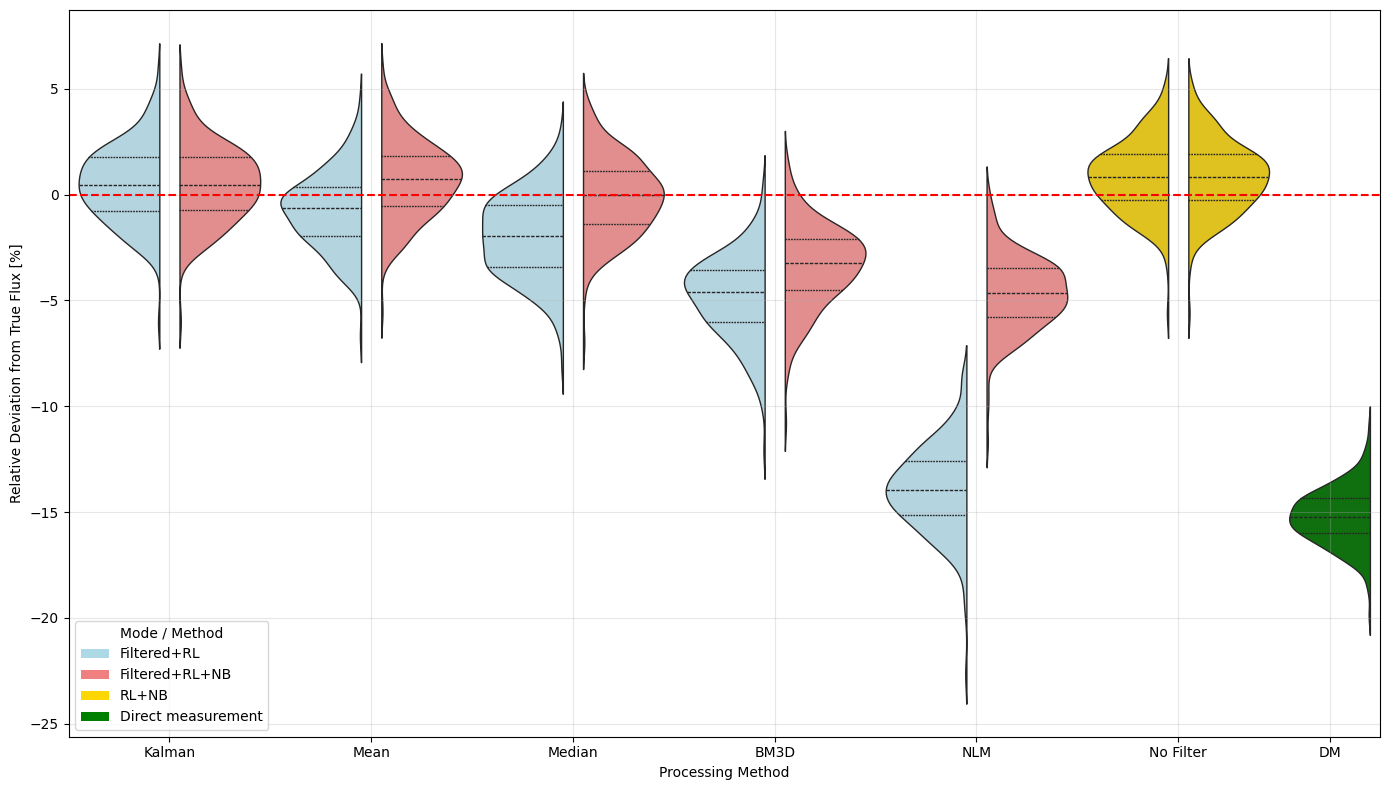}
\caption{Relative deviations of aperture photometry fluxes for Mock-II after 200 independent realizations. The true flux is defined as the flux within the red circular aperture in the ground-truth image, with measured fluxes extracted from the identical aperture in each deconvolved image. RL deconvolution algorithms incorporating noise backfilling (red) consistently outperform their nonbackfilling counterparts (blue), with the most pronounced differences observed for the median and NLM filtering methods. For reference, the green bar shows the direct aperture photometry result on the raw mock exposure image, which yields a flux approximately 15\% lower than the true value.}
\label{fig5}
\end{figure*}

We took the flux within the red aperture in the ground-truth image as the true flux, and the flux within the same aperture in each deconvolved image as the measured quantity. After 200 mock realizations, the relative deviations of the aperture photometry are shown in Figure \ref{fig5}. Similar to the isolated point-source case, RL algorithms with noise backfilling (red) generally outperform those without noise backfilling (blue); the most striking contrast occurs for the median and NLM filters. For comparison, we directly measured the aperture flux on the mock exposure image using the same aperture; the result is approximately 15\% lower than the true flux (green bar).

To obtain a broader comparison, we varied the input flux over three levels (2000, 6000, 18000 counts), the background over two levels (20 and 50 counts), and the S\'ersic index over two values ($n=4.0$ and $n=1.0$), performing multiple rounds of simulations. The results are presented in Table \ref{tab:rmd_comparison_mock2} ($n=4.0$) and Table \ref{tab:rmd_comparison_mock3} ($n=1.0$). From these two tables, it is clear that Kalman+RL generally has good flux conservation, and Kalman+RL+NB (with noise backfilling) is the best.

\begin{table*}
  \centering
  \caption{RMD (in percent) for S\'ersic index $n=4.0$. ``Overall mean'' is the average over different conditions for the same algorithm.}
  \label{tab:rmd_comparison_mock2}
  \resizebox{\textwidth}{!}{
  \begin{tabular}{lccccccc}
    \toprule
    RMD in percent & F2kB20 & F6kB20 & F18kB20 & F2kB50 & F6kB50 & F18kB50 & Overall mean \\
    \midrule
    Direct measurement & 15.51 & 15.29 & 15.28 & 15.12 & 15.21 & 15.29 & 15.28 \\
    RL+NB              & 1.01  & 0.57  & 0.40  & 1.24  & 0.85  & 0.37  & 0.74 \\
    Kalman+RL          & 0.17  & 0.42  & 0.41  & 0.44  & 0.50  & 0.33  & 0.38 \\
    {\bf Kalman+RL+NB}       & 0.18  & 0.42  & 0.41  & 0.41  & 0.49  & 0.33  & {\bf 0.37} \\
    Mean+RL            & 1.61  & 0.50  & 0.08  & 1.84  & 0.75  & 0.16  & 0.82 \\
    Mean+RL+NB         & 0.19  & 0.95  & 1.51  & 0.39  & 0.69  & 1.27  & 0.83 \\
    Median+RL          & 3.69  & 2.06  & 1.60  & 3.62  & 1.95  & 1.55  & 2.41 \\
    Median+RL+NB       & 1.01  & 0.00  & 0.38  & 0.91  & 0.09  & 0.35  & 0.46 \\
    BM3D+RL            & 9.85  & 3.14  & 0.86  & 14.41 & 4.85  & 1.52  & 5.77 \\
    BM3D+RL+NB         & 5.07  & 2.46  & 0.71  & 4.95  & 3.36  & 1.28  & 2.97 \\
    NLM+RL             & 67.63 & 4.88  & 0.26  & 77.73 & 13.93 & 0.74  & 27.53 \\
    NLM+RL+NB          & 9.28  & 1.87  & 0.27  & 11.11 & 4.69  & 0.28  & 4.58 \\
    \bottomrule
  \end{tabular}
  }
\end{table*}

\begin{table*}
  \centering
  \caption{RMD (in percent) for S\'ersic index $n=1.0$. The algorithm with the best flux conservation is shown in bold.}
  \label{tab:rmd_comparison_mock3}
  \resizebox{\textwidth}{!}{
  \begin{tabular}{lccccccc}
    \toprule
    RMD in percent & F2kB20 & F6kB20 & F18kB20 & F2kB50 & F6kB50 & F18kB50 & Overall mean \\
    \midrule
    Direct measurement & 20.35 & 20.12 & 20.10 & 19.95 & 20.02 & 20.08 & 20.10 \\
    RL+NB              & 0.94  & 0.49  & 0.04  & 0.67  & 0.67  & 0.26  & 0.51 \\
    Kalman+RL          & 0.72  & 0.01  & 0.22  & 0.57  & 0.13  & 0.09  & 0.29 \\
    {\bf Kalman+RL+NB}       & 0.69  & 0.01  & 0.22  & 0.56  & 0.13  & 0.09  & {\bf 0.28} \\
    Mean+RL            & 3.00  & 1.07  & 0.25  & 3.35  & 1.64  & 0.53  & 1.64 \\
    Mean+RL+NB         & 1.70  & 0.25  & 1.06  & 2.05  & 0.32  & 0.78  & 1.03 \\
    Median+RL          & 4.99  & 2.48  & 1.54  & 5.15  & 2.71  & 1.58  & 3.08 \\
    Median+RL+NB       & 2.65  & 0.73  & 0.06  & 2.71  & 1.12  & 0.06  & 1.22 \\
    BM3D+RL            & 10.20 & 3.12  & 0.71  & 14.68 & 5.06  & 1.37  & 5.86 \\
    BM3D+RL+NB         & 6.05  & 2.47  & 0.55  & 6.64  & 3.73  & 1.13  & 3.43 \\
    NLM+RL             & 69.11 & 7.82  & 0.10  & 78.39 & 20.50 & 1.54  & 29.58 \\
    NLM+RL+NB          & 13.88 & 3.60  & 0.07  & 16.41 & 7.70  & 0.79  & 7.08 \\
    \bottomrule
  \end{tabular}
  }
\end{table*}

\subsection{Mock-III: Crowded Stellar Fields}

In crowded stellar fields, PSF photometry has traditionally been the method of choice because it effectively decouples blended fluxes. However, conventional PSF photometry (e.g., the \texttt{Photutils} package, Version 2.3.0, \citealt{Bradley2023}) struggles to detect low-SNR sources, leading to missed detections and consequently affecting photometric accuracy. Given that in Mock-I we observed that the RL+NB algorithm produces a high plateau in the curve of growth near the true flux (e.g., the A14 region), we can potentially exploit this property to achieve high-precision aperture photometry on PSF-deconvolved images of crowded fields. Moreover, PSF-deconvolved images are more favorable for detecting faint sources. For example, \cite{Wang2025} used the upsampling and PSF deconvolution coaddition (UPDC; \citealp{Wang2022}) algorithm, which combines antialiasing and PSF deconvolution stacking, to detect 50\% more faint sources near the SMACS J0723 galaxy cluster in JWST images than the Drizzle stacking algorithm, and released the UPdec-Webb dataset. Using this dataset, \cite{Deng2025} discovered an extremely low-mass strong gravitational lensing system in the same region. Similarly, \cite{Zhang2025} used an antialiasing algorithm (fiDrizzle-MU, essentially a deconvolution of aliasing effects) to enhance image SNR and improve detection rates, discovering a strongly lensed quasar system in the SMACS J0723 field whose multiple images were below the detection threshold before deconvolution.

\begin{figure*}
\centering
\includegraphics[width=0.9\textwidth]{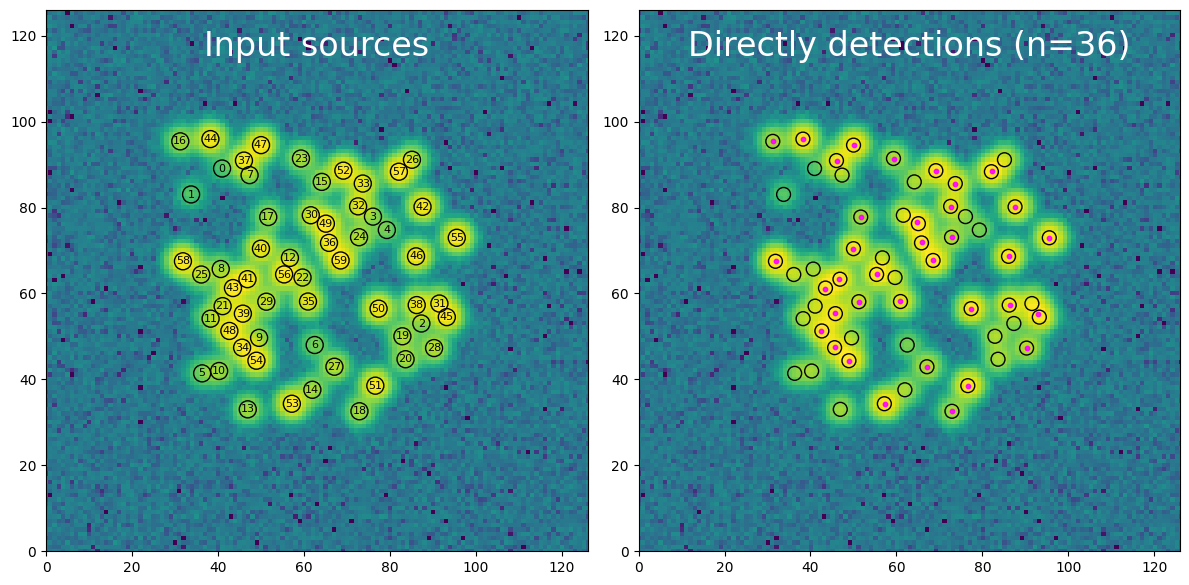}
\caption{Mock-III crowded stellar field (CSST/MCI simulator): 60 point sources (1000--10,000 counts, minimum 3.5 pixel separation) in a $63\times63$ pixel central region of a $127\times127$ pixel frame (left). Mock exposure is made by convolving with a Gaussian PSF ($\sigma = 2.0$ pixels, FWHM = 4.71 pixels), adding a 10 count background and Poissonization (left). \texttt{Photutils} detects only 36 sources (right), missing 24 faint or blended ones with overestimated fluxes; red dots mark detections; and tag-45 has a large center offset due to blending.}
  \label{fig6}
\end{figure*}

To simulate a crowded stellar field, we constructed Mock-III using the simulator dedicated to the Multi-channel Imager on board the Chinese Survey Space Telescope (CSST-MCI; \citealp{Yan2026a,Yan2026b}). Within the central $63\times63$ pixel area of a $127\times127$ pixel image, we placed 60 point sources with a minimum angular separation of 3.5 pixels to represent a dense stellar field. The input fluxes of these point sources were randomly drawn from 1000 to 10000 counts, and we labeled them in order of increasing flux, as shown in the left panel of Figure \ref{fig6}. Mock-III then convolved the sky image with a Gaussian PSF\footnote{For a more representative outcome, we did not employ the CSST PSF in our simulations; instead, we adopted a generic Gaussian PSF.} of $\sigma=2.0$ pixels (FWHM = 4.71 pixels), added a background of 10 counts, and finally Poissonized to obtain a mock exposure image (left panel of Figure \ref{fig6}). If we directly perform source detection and PSF photometry on the mock exposure using \texttt{Photutils}, we obtain 36 detections (the number in parentheses) and correspondingly somewhat overestimated photometric values; 24 faint or severely blended sources are missed, as shown in the right panel of Figure \ref{fig6}. Sources marked with red dots within black circles indicate those detected by \texttt{Photutils} and for which PSF photometry was performed; the red dot center is the detected source center. It can be seen that source tag-45, due to blending with surrounding sources, has its detected center significantly offset from its true center (the center of the black circle).

\begin{figure*}
\centering
\includegraphics[width=0.9\textwidth]{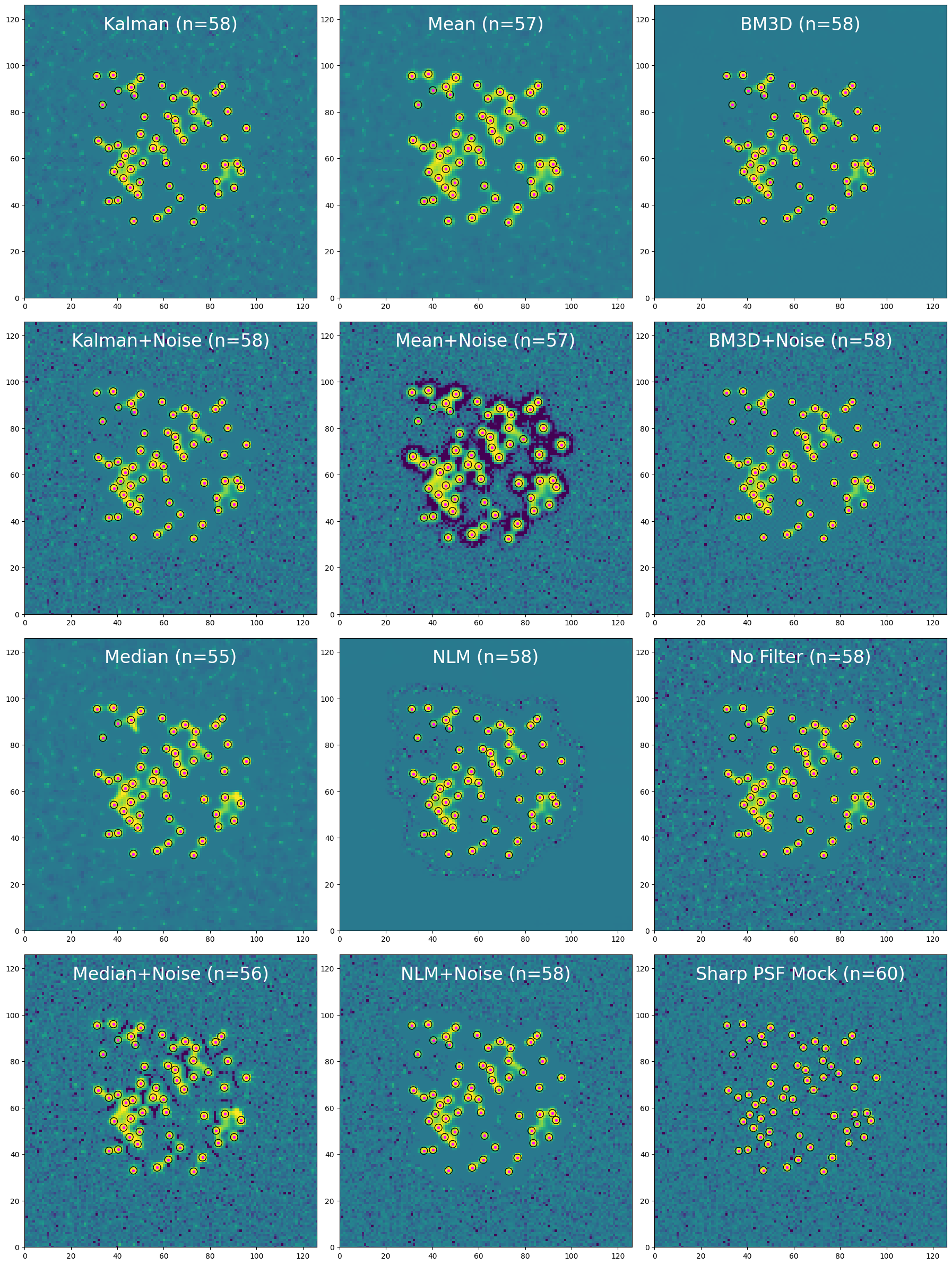}
\caption{Performance of six algorithms on Mock-III crowded stellar field (results in Table \ref{tab:matched_rmd}). All use 200 RL deconvolution iterations, uniform source detection, and 2-pixel radius aperture photometry; parenthetical numbers = detected sources. A sharp Gaussian PSF ($\sigma=0.5$) sky image (bottom row, third panel) recovers all 60 sources via \texttt{Photutils}. Mean/Median filtering oversmooths flux peaks (source loss) and creates bright-source troughs post noise backfilling. Four algorithms detect 58 sources (missing faint/blended tag-2 and tag-3); isolated faint tag-1 is enhanced by deconvolution. BM3D/NLM efficiently smooth background noise outside the crowded field.}
 \label{fig7}
\end{figure*}

Next, we applied the six algorithms to the mock exposure of the crowded field, and their performance is shown in Figure \ref{fig7} and Table \ref{tab:matched_rmd}. For all tested algorithms, RL deconvolution was performed with 200 iterations, followed by source detection under the same conditions, and aperture photometry using a fixed aperture radius of 2 pixels. The numbers in parentheses indicate the number of detected sources. As a reference, we convolved the sky image with a sharp Gaussian PSF ($\sigma=0.5$), added noise, and obtained the third image in the bottom row of Figure \ref{fig7}. We then performed source detection using \texttt{Photutils}, which recovered all 60 sources.

From Figure \ref{fig7}, we observe that Mean and Median filtering tend to oversmooth the flux peaks, causing some sources to be missed. Moreover, the filtered-out noise from these filters contains negative values with larger absolute magnitudes around bright sources, leading to pronounced troughs around bright sources after noise backfilling. The remaining methods (RL+NB, Kalman+RL, Kalman+RL+NB, BM3D+RL, BM3D+RL+NB, NLM+RL, and NLM+RL+NB) all detected 58 sources; they missed only the very faint and heavily blended tag-2 and tag-3. The faintest star (tag-1), although faint, is relatively isolated and thus becomes clearly prominent after PSF deconvolution. BM3D and NLM smooth the background noise very effectively in regions far from the crowded field.

Compared to the sharp-PSF mock, RL-deconvolved images exhibit a ``bridging'' phenomenon between two close sources: a flux bridge appears between them, becoming more prominent as the sources are closer. We suspect this is due to the non-linear nature of the RL algorithm; numerical simulations suggest that more iterations are required to break these bridges. Additionally, none of the RL deconvolution algorithms (with 200 iterations) produced spurious sources due to noise amplification.

\begin{table*}
  \centering
  \caption{Photometry Table for the Crowded Stellar Field}
  \label{tab:matched_rmd}
  \resizebox{\textwidth}{!}{
  \begin{tabular}{lcccc}
    \toprule
    & \multicolumn{2}{c}{All detections} & \multicolumn{2}{c}{High flux} \\
    \cmidrule(lr){2-3} \cmidrule(lr){4-5}
    & Matched Sources & relative mean deviation & Matched Sources & relative mean deviation \\
    \midrule
    Photutils PSF photometry & 36 & 4.60\% & 36 & 4.60\% \\
    RL+NB                     & 58 & 5.56\% & 44 & 4.39\% \\
    Kalman+RL                 & 58 & 5.34\% & 44 & 4.34\% \\
    {\bf Kalman+RL+NB}              & 58 & {\bf 5.02\%} & 44 & {\bf 3.97\%} \\
    Mean+RL                   & 57 & 8.26\% & 43 & 7.92\% \\
    Mean+RL+NB                & 57 & 6.01\% & 43 & 5.11\% \\
    BM3D+RL                   & 58 & 5.58\% & 44 & 4.45\% \\
    BM3D+RL+NB                & 58 & 5.45\% & 44 & 4.34\% \\
    Median+RL                 & 55 & 8.25\% & 42 & 7.56\% \\
    Median+RL+NB              & 56 & 6.29\% & 42 & 5.30\% \\
    NLM+RL                    & 58 & 5.57\% & 44 & 4.40\% \\
    NLM+RL+NB                 & 58 & 5.22\% & 44 & 4.14\% \\
    \bottomrule
  \end{tabular}
  }
\end{table*}

From Table \ref{tab:matched_rmd}, we note that most algorithms achieve aperture photometry errors of approximately 5\% on the RL-deconvolved crowded field. Because \texttt{Photutils} PSF photometry on the mock exposure missed 24 faint/blended sources, the PSF photometry of the detected sources is biased high (contaminated by the missing sources). The best performance is again achieved by the Kalman+RL+NB combination, with an aperture photometry error of 5\% for all 58 detected sources and below 4\% for high-flux sources.
In our mock tests with isolated stars, close extended sources, and crowded fields, the flux errors for RL+NB (with noise backfilling) and Kalman+RL+NB (filtering + noise backfilling) never exceeded 6\%; specifically, the maximum error for the latter was 5.02\% (in the crowded field).

\begin{figure*}
\centering
\includegraphics[width=0.9\textwidth]{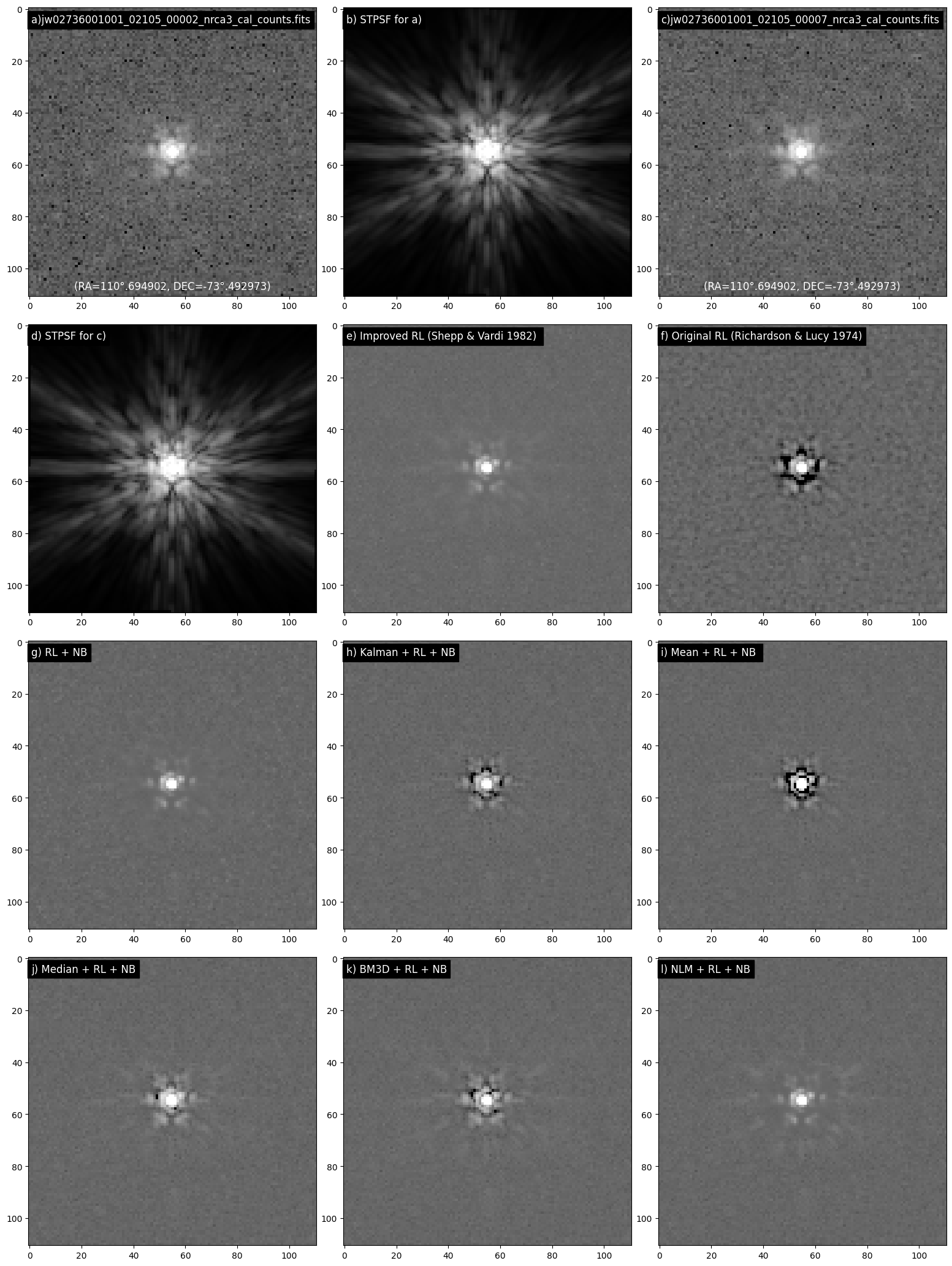}
\caption{Performance of eight methods on Observation-I (JWST/NIRCam F200W band). Panels $({\bold a})$ -- $({\bold d})$ display the exposures and STPSF characteristics: $({\bold a})$ raw exposure image in ADU; $({\bold b})$ logarithmic scale highlighting extended wings; $({\bold c})$ surface plot of the source; $({\bold d})$ simulated STPSF at the corresponding detector position. Panels $({\bold e})$ -- $({\bold l})$ show the stacked results after 10 RL iterations using different frameworks. Notably, the original RL $({\bold f})$ exhibits the highest noise level and ringing artifacts, while the FiDNoBa variants, panels $({\bold e})$, $({\bold j})$, $({\bold k})$, and $({\bold l})$ preserve diffraction spikes while suppressing background fluctuations.  Panels $({\bold e})$ -- $({\bold l})$ share the same color bar.}
 \label{fig8}
\end{figure*}

\begin{figure*}
\centering
\includegraphics[width=0.9\textwidth]{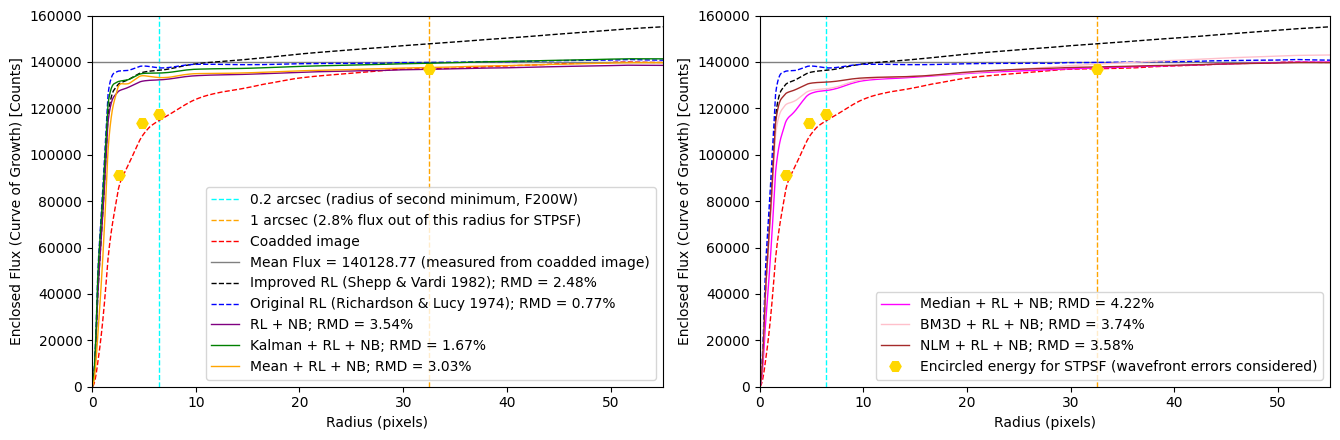}
\caption{Encircled energy curves (curves of growth) for the isolated star in Observation-I. The solid gray line represents the mean flux of nine exposures used as the ground-truth reference. Gold hexagons denote the the theoretical energy distribution of the STPSF (including wave front errors). Both original RL (pink dashed) and Kalman+RL+NB (purple solid) reach a stable plateau near the reference flux between $0.2^{\prime\prime}$ and $1.0^{\prime\prime}$. The original RL shows unexpectedly high stability in this specific case, likely due to the  ``compact core, extended wings''  profile of the JWST PSF.}
 \label{fig9}
\end{figure*}

\section{Application to Real Observational Data Processing}
\label{sec5}

\subsection{Observation-I: Application to JWST Data Processing}
The data used for this test are located in the parallel field of the SMACS J0723 galaxy cluster, i.e., images taken by the Module A detector of JWST/NIRCam. The data were downloaded as Stage-2 data (with the \_cal.fits suffix) from the Mikulski Archive for Space Telescopes database (i.e., Stage-2: SMACS J0723). The data were produced by the data processing software (SDP\_VER 2023\_4a), the calibration software (CAL\_VER 1.13.3; \citealt{Bushouse2019}), and the calibration reference data system (CRDS\_VER 11.17.14 and CRDS\_CTX jwst\_1223.pmap). We selected a relatively isolated star in the F200W band ($RA = 110^\circ.694902, Dec = -73^\circ.492973$) as the target for this test; the star is fully covered by nine exposures. Additionally, we used \texttt{STPSF} (Version 2.2.0) from the official \texttt{JWST pipeline} -- formerly known as WebbPSF -- to generate the PSF corresponding to each exposure at that position.

First, we converted the image units from MJy sr$^{-1}$ to analog-to-digital units (ADUs). Second, outliers such as cosmic rays, NaNs, and bad pixels in the multiple exposures were filled using statistical values from pixels at the same position on the other exposures (for details, refer to \citealt{Wang2024,Wang2025}). Finally, we performed PSF deconvolution on the nine exposure images using the STPSF corresponding to each exposure and the deconvolution method described in this paper. The procedure is similar to that for Mock-I, except that the mock data are replaced here with real observational data.

Figure \ref{fig8} presents the processing results for Observation-I using eight methods (deconvolution iterated 10 times, following \citealt{Wang2025}). From panels $({\bold a})$ to $({\bold d})$, it can be seen that the JWST PSF is very extended in the outer regions. Moreover, the observational data reveal that most of the outer structure is submerged in noise, indicating that the STPSF possesses a very sharp core; the radius of the first Airy minimum encloses 71\% of the total flux (i.e.,``compact core, extended wings''; see the JWST official document ``JWST-STScI-001157.pdf''\footnote{\url{https://www.stsci.edu/files/live/sites/www/files/home/jwst/documentation/technical-documents/_documents/JWST-STScI-001157.pdf}}). Panels $({\bold e})$ through $({\bold l})$ all show the stacked output images after PSF deconvolution of the nine exposures. Panels $({\bold e})$, $({\bold j})$, $({\bold k})$, and $({\bold l})$ exhibit relatively prominent diffraction spikes in the outer regions, whereas panel $({\bold f})$, corresponding to the original RL method, displays the largest noise level and the most significant ringing artifacts. Readers should note that the ringing effect referred to in this work is induced by PSF deconvolution, not by analogous structures inherent to the original PSF, e.g. the minima and maxima in the Airy pattern of the PSF.

Figure \ref{fig9} shows the encircled-energy curves as a function of radius. Since the true input flux is not known a priori, we adopt the mean flux of the nine observations as a reference value, shown as a solid gray line. In addition, the STPSF profile including wave front-error effects is marked with gold hexagons (here the flux percentages from the JWST official document ``JWST-STScI-001157.pdf" have been multiplied by the mean flux). It can be seen that in the interval from $0\farcs2$ (corresponding to the radius of the second Airy minimum, i.e., the hexagonal core region) to $1\farcs0$, the results produced by both the original RL and Kalman+RL+NB exhibit a plateau very close to the mean flux--this is an ideal region for performing aperture photometry. Surprisingly, the original RL method yields the smallest RMD (0.77\%), and its oscillations after first crossing the mean flux line are smaller than those shown in Mock-I. We speculate that this phenomenon arises from the ``compact core, extended wings'' characteristic of the STPSF. In the Observation-I test, the aperture photometry accuracy provided by Kalman+RL+NB is second only to that of the original RL method.

Furthermore, considering that the central region of the STPSF is usually sharper than the observed data, we conducted the same test using the data-driven\footnote{The central region of The central region of HybPSF is fitted using smoothed principal component analysis based on the stars within the field of view, while the peripheral region is derived from the STPSF.} HybPSF \citep{Nie2024} and obtained similar conclusions. In Appendix~\ref{APD}, we conduct an STPSF-based simulation and obtain similar conclusions.

\begin{figure*}
\centering
\includegraphics[width=0.9\textwidth]{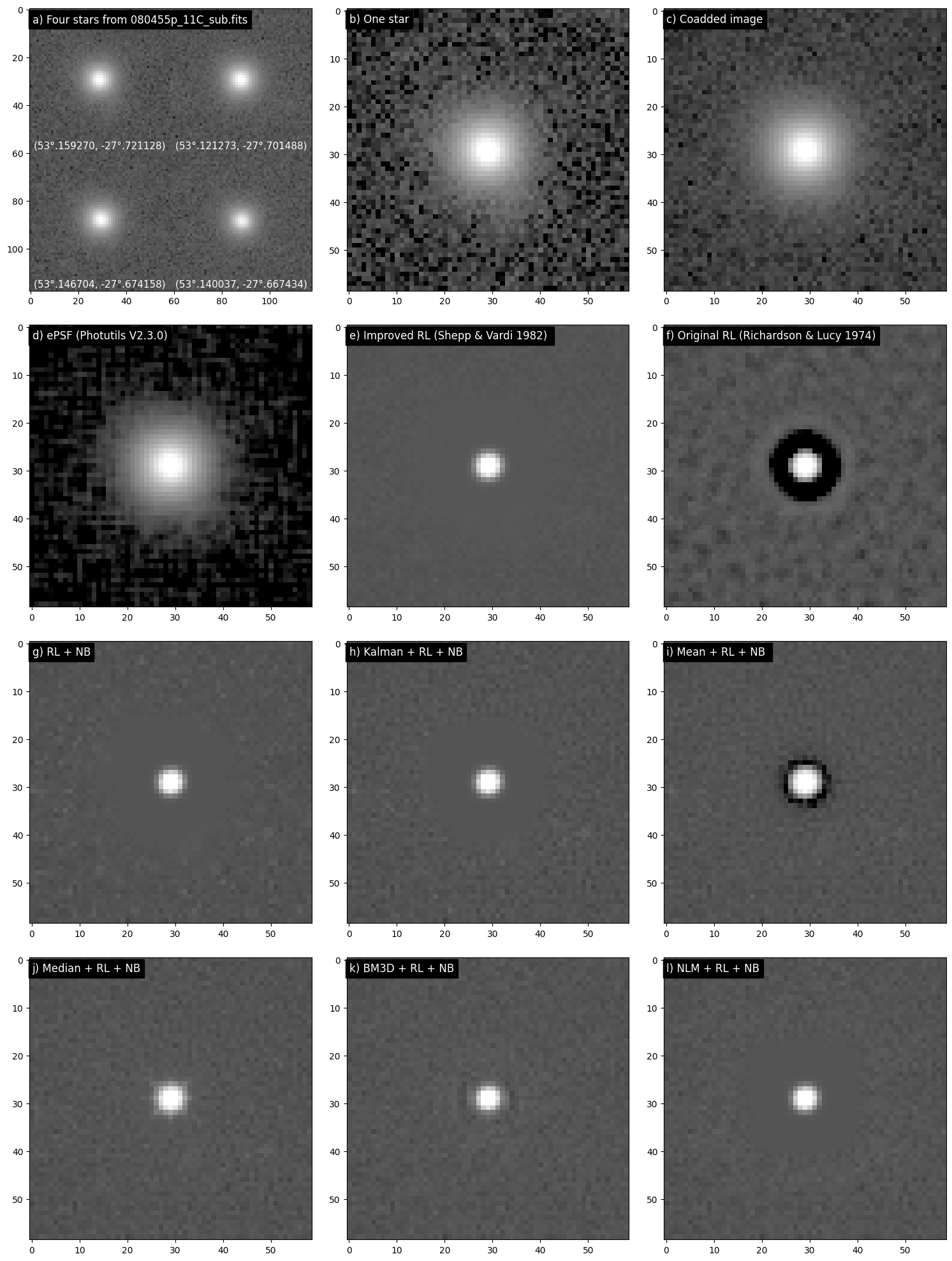}
\caption{Performance of eight methods on Observation-II (VST OmegaCAM $r$-band). Panel $({\bold a})$ identifies the four target stars used for ePSF construction; $({\bold b})$ shows a close-up of a single star; $({\bold c})$ is the average stack of the four sources; $({\bold d})$ displays the effective PSF (ePSF) generated via \texttt{Photutils}. Panels $({\bold e})$ -- $({\bold l})$ show the deconvolution results after 50 iterations. Severe ringing artifacts are visible in the original RL $({\bold f})$ and Mean+RL+NB $({\bold h})$ results, while the Median+RL+NB $({\bold i})$ method introduces spurious star-shaped morphologies. Kalman+RL+NB $({\bold g})$ maintains the most physically plausible source profile.  Panels $({\bold e})$ -- $({\bold l})$ share the same color bar.}
 \label{fig10}
\end{figure*}

\begin{figure*}
\centering
\includegraphics[width=0.9\textwidth]{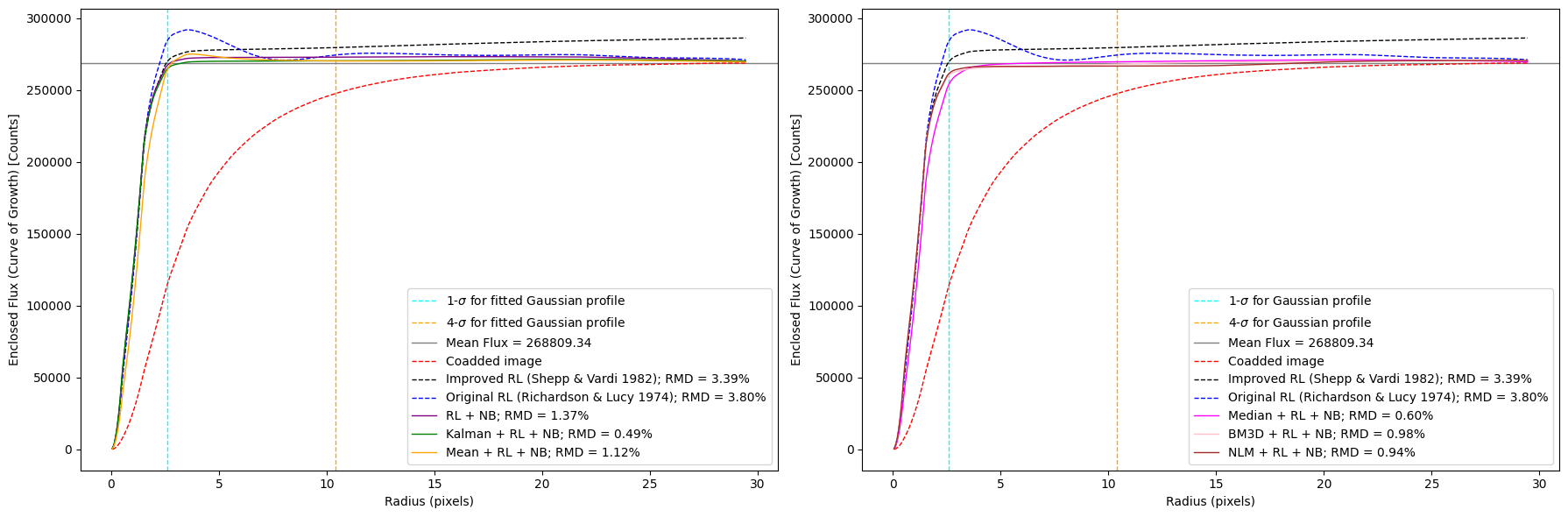}
\caption{Encircled energy curves for the VST stellar targets in Observation-II. The gray line indicates the mean measured flux of the four stars. Vertical dashed lines mark the $1\sigma$ and $4\sigma$ positions of a Gaussian fit to the ePSF. Within the A14 region, the Kalman+RL+NB method (purple solid) demonstrates the most accurate and stable flux recovery, significantly outperforming the original RL and improved RL, which exhibit violent oscillations and systematic overestimation, respectively.}
 \label{fig11}
\end{figure*}

\subsection{Observation-II: Application to VST Data Processing}
We applied the algorithm described in this paper to process data from a ground-based telescope, specifically $r$-band VOICE-CDFS-1 exposures ($080455p\_11C.sub.fits$) obtained with the OmegaCAM instrument on the VST of the European Southern Observatory.

First, we restored the original background based on the background Poisson noise in the image $080455p\_11C.sub.fits$.
Second, we selected four stars of comparable brightness located close together in the field of view as point-source targets. Using these stars, we constructed an effective PSF (ePSF) using the ePSF tools in \texttt{Photutils}.
Finally, assuming that these four stars share the same ePSF, we performed PSF deconvolution on the four stars using that ePSF and the deconvolution algorithm described in this paper. The procedure is similar to that for Mock-I, except that the mock data are replaced here with VST observational data.

Figure \ref{fig10} shows the processing results for Observation-II using eight methods (deconvolution iterated 50 times). Panel $({\bold a})$ lists all four stars and their celestial positions (the R.A. and decl. values in parentheses are in degrees); $({\bold b})$ is a zoomed-in view of one of the stars; $({\bold c})$ shows the stacked image of the four stars; and and (d) displays the ePSF extracted from these four stars using \texttt{Photutils}. Panels $({\bold e})$ through $({\bold l})$ are all stacked output images after PSF deconvolution of the four stars. It is clearly evident that the results produced by the original RL and Mean+RL+NB methods exhibit strong ringing artifacts, whereas the image output by the Median+RL+NB algorithm displays a spurious star-shaped structure.

Figure \ref{fig11} shows the encircled-energy curves as a function of radius. Since the true input flux is not known a priori, we adopt the mean flux of the four stars' observations as a reference value, shown as a solid gray line. For ease of comparison, we fitted the ePSF with a Gaussian function and marked the $1\sigma$ and $4\sigma$ positions with vertical lines in the figure. Between these two positions, Kalman+RL+NB exhibits a plateau closest to the mean flux reference line. Although Median+RL+NB is the next best, it performs better in the latter part of the interval. Moreover, the original RL once again displays violent oscillations, which is consistent with its behavior in Mock-I. It is evident that for a PSF with an extended core, this characteristic is unavoidable with the original RL method. Likewise, the improved RL method continues to show systematically overestimated flux measurements, which is also consistent with the results from Mock-I.

In this section, we applied the FiDNoBa and traditional algorithms to observational data from JWST/NIRCam (a space telescope, Observation-I) and VST (a ground-based telescope, Observation-II). We find that Kalman+RL+NB still exhibits excellent performance in terms of aperture photometry accuracy. In the case of Observation-I, its performance is second only to that of the original RL algorithm. Additionally, our privately conducted simulation tests based on STPSF also support the validity of the processing for Observation-I. The ``compact core, extended wing'' characteristic of the STPSF also leads, in simulations, to higher aperture photometry accuracy for images processed with the original RL algorithm.

\section{Conclusions}
\label{sec6}

This study presents significant enhancements to the classic Richardson--Lucy (RL) deconvolution algorithm, aimed at improving its photometric accuracy for astronomical applications. We successfully addressed the algorithm's inherent limitations regarding negative-valued images and noise amplification through two key innovations.

First, the introduction of a negative backfilling technique allows the RL algorithm to process background-subtracted images containing negative pixels while enforcing strict flux conservation. This method alone demonstrated a substantial improvement over the original and improved RL algorithms, reducing the relative mean deviation (RMD) in flux by a factor of $\sim3$ for isolated point sources.

Second, we developed the more comprehensive Filtered Image Deconvolution and (Filtered-out) Noise Backfilling (FiDNoBa) algorithm. This approach strategically combines low-pass filtering with RL deconvolution and noise backfilling. By applying a filter (e.g., Kalman, mean, median, BM3D, NLM) to suppress high-frequency noise prior to deconvolution and then restoring the filtered-out noise post-deconvolution, FiDNoBa effectively mitigates the noise amplification characteristic of RL iterations while preserving fine spatial details and photometric integrity.

Extensive numerical simulations across three scenarios---isolated point sources (Mock-I), closely blended S\'ersic galaxies (Mock-II), and dense crowded stellar fields (Mock-III) -- provided a comprehensive evaluation. The results consistently highlight the superior performance of the Kalman+RL+NB combination. This specific configuration achieved the lowest RMD across nearly all test conditions: $\sim0.05$ -- $0.85\%$ for isolated sources, $\sim0.28$ -- $0.37\%$ for blended galaxies, and $\sim5.02\%$ for all detected sources in a crowded field. Notably, in the crowded field test, our method recovered 58 sources compared to 36 found by traditional PSF-fitting photometry (using \texttt{Photutils}), demonstrating its enhanced capability for source detection in blended regimes. The resulting deconvolved images exhibit a characteristic ``high plateau'' in their curve of growth near the true flux value, which is highly beneficial for robust aperture photometry.

Finally, the Kalman+RL+NB variant of FiDNoBa provides robust and accurate photometric performance across both space-based (JWST) and ground-based (VST) observations. While original RL performed surprisingly well on JWST data due to the unique shape of the STPSF (see also a mock test for JWST/NIRCam in the appendix), Kalman+RL+NB offers a more generally reliable solution, particularly for extended PSFs where RL alone is prone to ringing and bias.

In summary, the FiDNoBa framework, especially when employing Kalman filtering, provides a powerful and reliable tool for astronomical image deconvolution. It delivers high photometric accuracy, improves source detection in crowded areas, and is applicable to both point-like and extended sources. Future work will focus on optimizing filter selection for specific astronomical use cases, extending the method to multiwavelength data, and integrating it into automated data reduction pipelines for current and future survey telescopes.

\section*{Acknowledgments}
We are grateful for the valuable comments provided by the anonymous reviewer, which have greatly improved the presentation of this paper. In particular, the reviewer's attention to detail and insightful comments have profoundly enhanced the potential impact of this work, and we express our special thanks to the reviewer. This work is supported by the National Natural Science Foundation of China (Nos. 12503009, 12403007, 12573115 and 12533008), the China Manned Space Project (Nos. CMS-CSST-2021-A01, CMS-CSST-2021-A03, CMS-CSST-2021-A04, CMS-CSST-2021-A07, CMS-CSST-2025-A20, and CMS-CSST-2021-B01). X.P. acknowledges the support from the Strategic Priority Research Program of the Chinese Academy of Sciences (grants Nos. XDA0350201).

\appendix
\section{A Mock Test for JWST/NIRCam F200W}
\label{APD}

\begin{figure*}
\centering
\includegraphics[width=0.9\textwidth]{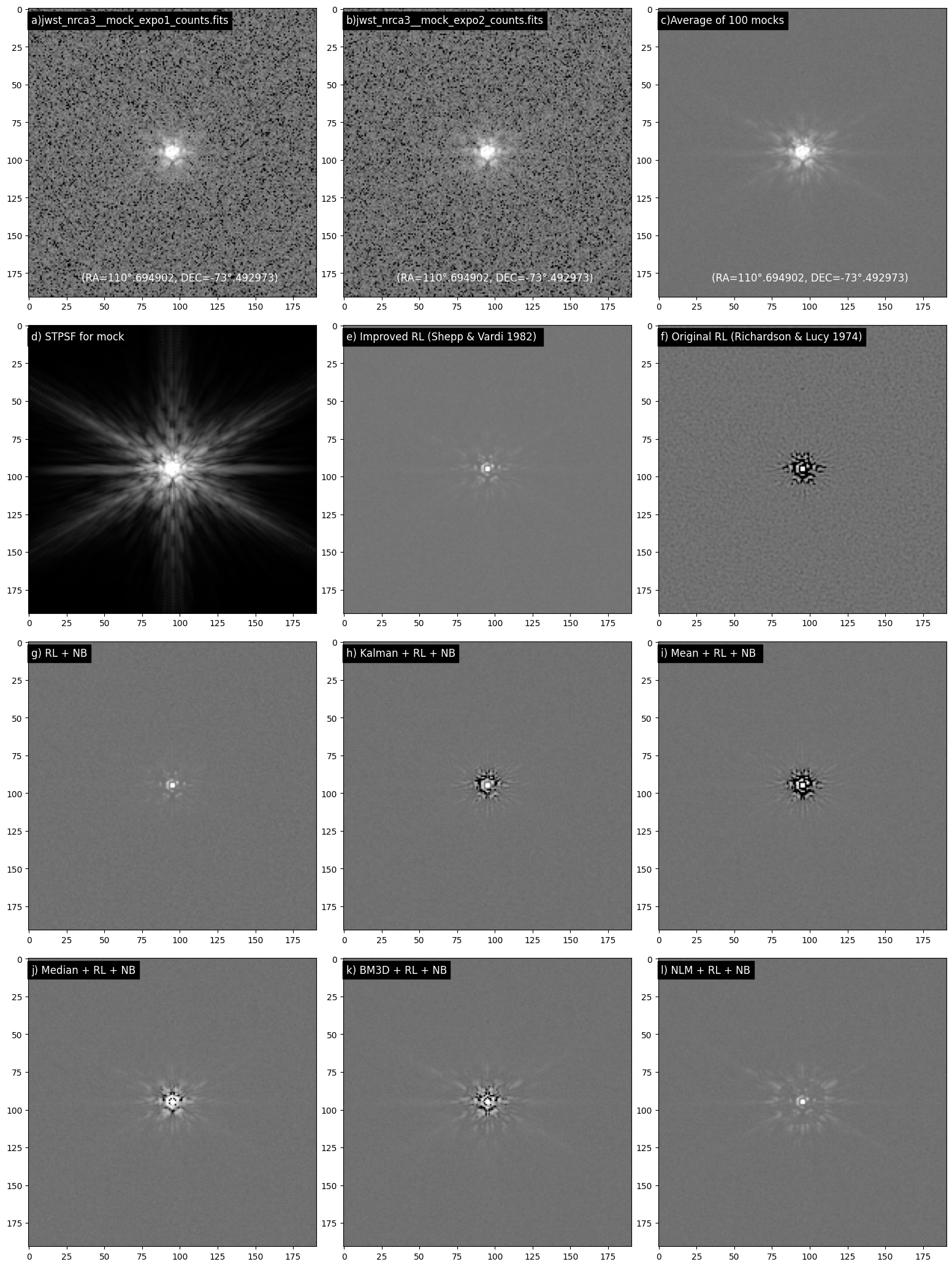}
\caption{Performance of eight methods on a JWST/NIRCam F200W mock test. Panels $({\bold a})$ -- $({\bold d})$ display the mock exposures and STPSF characteristics: $({\bold a})$ and $({\bold b})$ mock exposures in ADU; $({\bold c})$ a stack of 100 mock exposures; $({\bold d})$ simulated STPSF at the corresponding detector position. Panels $({\bold e})$ -- $({\bold l})$ show the stacked results after 10 RL iterations using different frameworks.  Panels $({\bold a})$ -- $({\bold c})$ and $({\bold e})$ -- $({\bold l})$ share the same color bar.}
 \label{figAP1}
\end{figure*}

\begin{figure*}
\centering
\includegraphics[width=0.9\textwidth]{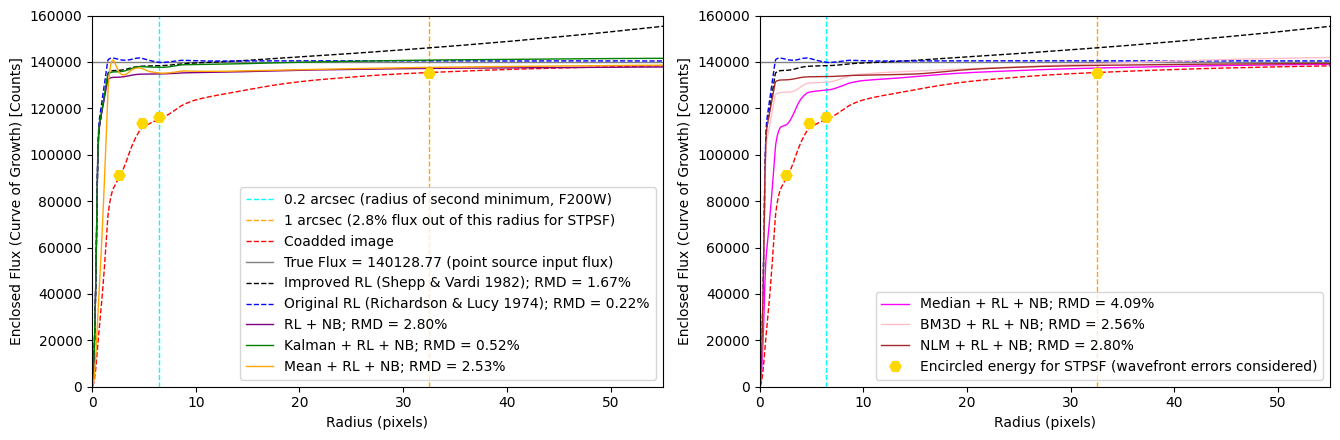}
\caption{Encircled energy curves for a simulated isolated star. The solid gray line represents the input flux of 100 exposures used as the ground-truth reference. Gold hexagons denote the the theoretical energy distribution of the STPSF (including wave front errors). Both original RL (pink dashed) and Kalman+RL+NB (purple solid) reach a stable plateau near the reference flux between $0.2^{\prime\prime}$ and $1.0^{\prime\prime}$. }
 \label{figAP2}
\end{figure*}

To further understand the behavior of the different deconvolution schemes for a JWST-like ``compact core, extended wings'' PSF, we performed an additional mock test based on an STPSF model for the NIRCam F200W band. The STPSF was generated at the detector position corresponding to the real target used in Observation-I. Specifically, we adopted the following configuration:
{\scriptsize
\begin{verbatim}
input_fits = "jw02736001001_02105_00007_nrca3_cal.fits"
detector   = "NRCA3"
ccd_x      = 1685
ccd_y      = 1330
fov_pixels = 255
osam       = 1
nlambda    = 10
normalize  = True
\end{verbatim}
}

The resulting STPSF is shown in panel $({\bold d})$ of Figure~\ref{figAP1}.

Using this PSF, we constructed a set of simulated point-source images. A point source with a total flux of $140128.77$ counts was first convolved with the STPSF, and a constant background of $62.31$ counts was then added. Both the source flux and the background level were measured from the real observation at the same location. Finally, we Poissonized the image using 100 different random seeds, without adding any additional instrumental effects and without applying any dithering pattern. The first two simulated exposures are shown in panels $({\bold a})$ and $({\bold b})$ of Figure~\ref{figAP1}. Panel $({\bold c})$ is a stack of 100 mock images. Each of the 100 mock images was then processed with the same set of image-processing methods used in the main text, and the processed images were coadded. The final stacked results after 10 RL iterations are displayed in panels $({\bold e})$--$({\bold l})$ of Figure~\ref{figAP1}. The corresponding encircled-energy curves, or curves of growth, are shown in Figure~\ref{figAP2}.

Combining Figures~\ref{figAP1} and \ref{figAP2}, we find that both the original RL and Kalman+RL+NB methods significantly suppress the extended wings and diffraction spikes of the STPSF. However, the encircled-energy curves of the original RL and Mean+RL+NB methods show oscillatory structures with alternating peaks and troughs, including regions where the radial derivative becomes negative. This behavior indicates the presence of ringing-like artifacts in the processed images. The Kalman+RL+NB, Median+RL+NB, and BM3D+RL+NB results also show visually similar radial structures, but these are mainly associated with the intrinsic STPSF morphology, such as the first, second, and third Airy minima. Unlike ringing artifacts, these Airy minima do not correspond to negative flux depressions.

The gold hexagonal markers in Figure~\ref{figAP2} show that the radial profile of the mock coadded image agrees well with the theoretical JWST PSF expectation from the JWST PSF documentation. This behavior differs from that of the real observation shown in Figure~\ref{fig9}, suggesting that the central profile of the observed star is less steep than that of the idealized STPSF simulation. This is consistent with the conclusion of \cite{Nie2024}. One possible reason is that the JWST dithering strategy introduces an additional pixelization-blurring effect in multiexposure observations, which can be described as a noise-spike kernel \citep{Wang2022}. In contrast, in the present mock test the point source is placed exactly at the center of a detector pixel, and no dithering is included. This difference is also reflected by comparing Figures~\ref{fig9} and \ref{figAP2}: for the same method, the curves in Figure~\ref{figAP2} approach a flat plateau more rapidly than those in Figure~\ref{fig9}.

If aperture photometry is performed for a point source using radii from $0.2^{\prime\prime}$, approximately the radius of the second Airy minimum, to $1.0^{\prime\prime}$, the original RL method provides the most reliable estimate in this mock test, with a mean absolute relative deviation of only $0.22\%$. The Kalman+RL+NB method ranks second, with a mean absolute relative deviation of $0.52\%$. Table~\ref{tab:deviation_comparison} summarizes the relative mean deviations of the different methods at $0.2^{\prime\prime}$ and $1.0^{\prime\prime}$ for both the real observation and the mock data.  As reported by \citet{Leist2024}, flux errors range from 9.19\% to 14.11\% (across the F560W to F1000W bands) when the RL algorithm is adopted for aperture photometry in JWST-MIRISim simulations. Intriguingly, the RL method employed in their work shows no suppression of the outer wings of the STPSF, see Figure 5 in \citep{Leist2024}, whereas this suppression effect is prominent in our proposed method.

We speculate that the unexpectedly good performance of the original RL method in this STPSF-based mock is caused by a partial cancellation effect: the angular ringing introduced by original RL may compensate for the residual contribution of the extended PSF wings after deconvolution, allowing the encircled-energy curve to approach the theoretical flux level rapidly and then oscillate around it. A similar effect is also visible in the Mean+RL+NB result. However, Mean+RL+NB modifies the central peak more strongly and does not suppress the outer wings as effectively as original RL, leading to a less accurate photometric result.

In summary, for a nearly undersampled JWST PSF such as NIRCam F200W, which is close to the critical sampling regime and has a compact core with extended wings, the original RL method is the most favorable for aperture photometry on deconvolved images, despite the presence of some ringing-like artifacts. Kalman+RL+NB provides the second-best performance in this specific case. By contrast, Kalman+RL+NB is more suitable for ground-based observations, where the PSF core is usually more extended, as demonstrated by the VST data in Section~\ref{sec5}.

 We emphasize that developing aperture photometry methods for PSF-deconvolved images is vital for bright extended-source photometry with JWST. Substantial flux leakage arises from the STPSF wings of such sources, making conventional extended-source models and point-source-PSF composite modeling ineffective.

\begin{table}
  \centering
  \caption{Relative Mean Deviation Comparison for a JWST Observation and Mock Data}
  \label{tab:deviation_comparison}
  \begin{tabular}{lcccc}
    \toprule
    & \multicolumn{2}{c}{Observation} & \multicolumn{2}{c}{Mock} \\
    \cmidrule(lr){2-3} \cmidrule(lr){4-5}
    Method & $0\farcs2$ & $1\farcs0$ & $0\farcs2$ & $1\farcs0$ \\
           & (\%) & (\%) & (\%) & (\%) \\
    \midrule
     {\bf original RL}   & {\bf 1.73} & {\bf 0.25} & {\bf 0.04} & {\bf 0.23} \\
    RL+NB         & 5.69 & 2.37 & 3.81 & 2.10 \\
    improved RL          & 2.81 & 5.47 & 1.31 & 4.23 \\
    Kalman+RL+NB     & 3.54 & 0.49 & 1.70 & 0.46 \\
    Mean+RL+NB       & 4.76 & 1.80 & 3.29 & 1.78 \\
    Median+RL+NB     & 9.05 & 1.96 & 8.89 & 1.98 \\
    BM3D+RL+NB       & 8.48 & 0.79 & 6.39 & 0.65 \\
    NLM+RL+NB        & 6.37 & 1.63 & 4.62 & 1.16 \\
    \bottomrule
  \end{tabular}
\end{table}

\section*{ORCID iDs}
Lei Wang \href{https://orcid.org/0000-0002-3779-9069}{https://orcid.org/0000-0002-3779-9069}\\
Xiyan Peng \href{https://orcid.org/0000-0002-3784-830X}{https://orcid.org/0000-0002-3784-830X}\\
Huanyuan Shan \href{https://orcid.org/0000-0001-8534-837X}{https://orcid.org/0000-0001-8534-837X}\\
Fengshan Liu \href{https://orcid.org/0009-0001-7105-2284}{https://orcid.org/0009-0001-7105-2284}\\
Zhenya Zheng \href{https://orcid.org/0000-0002-9634-2923}{https://orcid.org/0000-0002-9634-2923}

\label{lastpage}
\end{document}